# Identifying Fake Profiles in LinkedIn

Shalinda Adikari

National University of Singapore

shalinda@comp.nus.edu.sg

Kaushik Dutta

University of South Florida

duttak@usf.edu

Abstract

*As organizations increasingly rely on professionally oriented networks such as LinkedIn (the largest such social network) for building business connections, there is increasing value in having one's profile noticed within the network. As this value increases, so does the temptation to misuse the network for unethical purposes. Fake profiles have an adverse effect on the trustworthiness of the network as a whole, and can represent significant costs in time and effort in building a connection based on fake information. Unfortunately, fake profiles are difficult to identify. Approaches have been proposed for some social networks; however, these generally rely on data that are not publicly available for LinkedIn profiles. In this research, we identify the minimal set of profile data necessary for identifying fake profiles in LinkedIn, and propose an appropriate data mining approach for fake profile identification. We demonstrate that, even with limited profile data, our approach can identify fake profiles with 87% accuracy and 94% True Negative Rate, which is comparable to the results obtained based on larger data sets and more expansive profile information. Further, when compared to approaches using similar amounts and types of data, our method provides an improvement of approximately 14% accuracy.*

# 1  INTRODUCTION

In recent years, social networks have had a dramatic impact on human social interactions, changing the web into a social web where users and their communities are centres for online growth, commerce, and information sharing [1]. Social networks each offer unique value chains targeting different user segments. Users find old friends perusing Facebook, and receive fast updates and breaking news through Twitter. LinkedIn is designed to support professional communities, where users maintain a profile with a high degree of personal contacts, and search for contacts with desired skills. Measured by usage rates, Facebook is the most-visited social network with 800 million visitors per month; Twitter is the second most-visited social network with 250 million visitors per month. LinkedIn has the third-highest visit rate, with more than 200 million visitors per month, and serves as the world's largest professional network [2].

The surge of social networks' popularity, combined with the availability of large amounts of information posted by users, from email addresses to personal information and messages, make users attractive targets for malicious entities. Most attacks focus on retrieving user information without user consent. Typically, an attacker will create a fake profile, and then solicit a connection to the intended victim. If the target accepts the connection, the attacker then has access to the target's full profile, which would otherwise be accessible only to trusted contacts [3]. Further, each victim "connection" a fake profile can attract can serve to increase the profile's surface legitimacy, which can help to attract further connections.

According to Cloudmark estimates, between 20% and 40% of Facebook and Twitter accounts could be fake profiles [4]. Due to the high levels of user interaction and the millions of daily transactions, separating suspicious users from legitimate ones is increasingly difficult. Efforts to solicit user assistance in flagging fake profiles have not achieved the results expected [5]. Further, in networks with strict user privacy policies, little data is publicly available, making it difficult to distinguish between fake and legitimate profiles in a systematic way before deciding whether to trust a potential connection. In this work, we propose a method for differentiating between legitimate and fake profiles in a methodical manner, based on limited publicly available profile information in sites with restrictive privacy policies, such as LinkedIn.

## 1.1  Background

A typical social network profile consists of two main parts, static data and dynamic data. Static data refers to slowly changing or unchanging information that the user enters into the system, while dynamic data refers to information describing the user's activity on the social network. The set of static data typically includes a user's demographics and interests, while dynamic data relates to user activities, connections, and position in the social network [6].

A fake profile is a social network profile that maintains a forged identity purporting to be someone other than the person who created it, or containing fictitious personal information. Krombholz, Merkl [7] observed that user behavior on fake profiles is different from that of legitimate users. There are three significant ways of creating fake social network profiles. First, a legitimate user may fabricate the content in his own profile [7] in order to manipulate the appeal of his profile [8, 9]. Second, a user may clone the profile of another legitimate user [10, 11] by creating a similar profile in the same or another social network by copying the victim's profile and adding victim's friends into the new fake profile. Third, a user may create a profile with a fabricated identity [7] in order to attain victims' trust and confidence, and then collect their confidential information. Because the features of legitimate and fake profiles are very similar on the surface, it is difficult to distinguish between them without a reliable method for using the available data for differentiation purposes.

Figure 1 shows two sample fake profiles: the profile on the left contains data that does not make sense together, while the profile on the right shows two profiles with the same photo, but slightly different content.

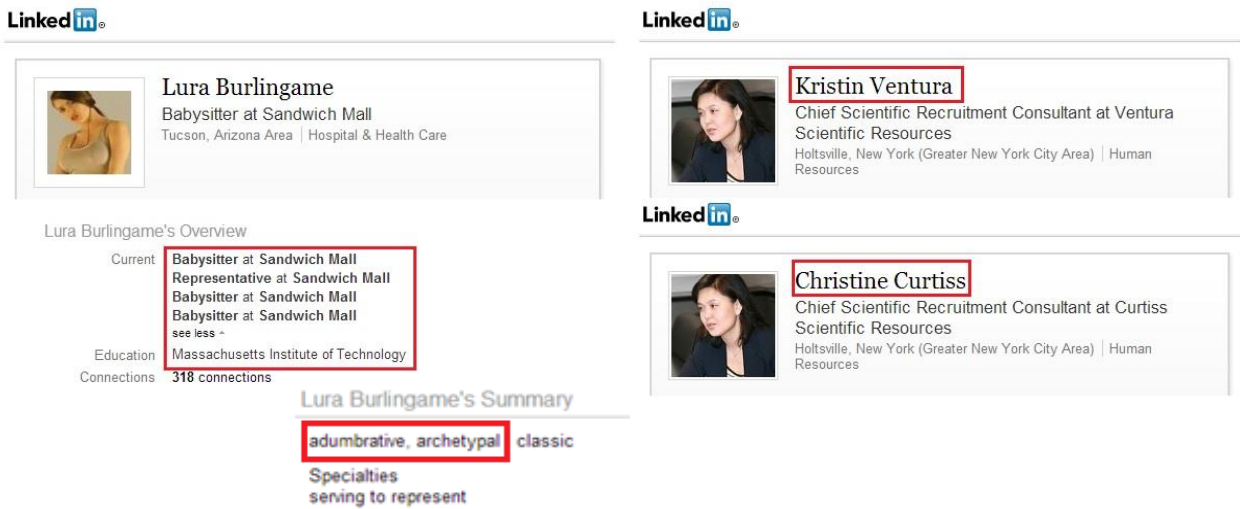

*Figure 1: Examples for LinkedIn fake profiles: on the left, a bot-generated profile [12]; on the right, two profiles with the same photo and different name [13]*

## 1.2 Problem Statement

The focus of this research is to develop a method for recognizing and differentiating legitimate profiles and fake profiles in LinkedIn. LinkedIn is the most prominent professional social network, supporting the management of users' digital resumes. In this network, users can list biographical data, professional history, and maintain a list of professional contacts. The concentration of professional information on a single platform has created a valuable repository of professional information of interest to a variety of stakeholders. For instance, there is a significant trend in using LinkedIn for recruiting and job searching – recruitment agents can select potential workers and users can identify and contact potential employers. These activities have actual monetary value, which has attracted the notice of malicious users and increased the rate of fake profile creation. For example, in early 2014, a group of hackers executed a Botnet attack and created thousands of fake LinkedIn profiles [14]. Currently, the identification of fake profiles in LinkedIn is limited to a manual reporting process, where a user can flag a suspected fake profile.

Virtually all current research approaches proposed for identifying fake profiles depend on both static and dynamic data, and are based on either Facebook or Twitter. Both of these social networks provide rich Application Programming Interfaces (APIs) to provide relevant, real-time, and up-to-date user information supporting the research requirements. The Facebook API [15] facilitates access to a wide variety of both static and dynamic profile information, including user activities, friends' activities, friends of friends, and basic biographical details (including age, birthday, profile status, relationship status, likes, groups details, etc.). Similarly, the Twitter API [16] provides twitter counts, followers, notifications, friends, and basic user details. In contrast, LinkedIn has much stricter privacy policies, with limited public visibility for static profile information, and no access to dynamic profile details. Due to these stringent privacy policies [17], it is difficult, if not impossible, to apply existing practical and theoretical approaches for fake profile detection. In this research, our goal is to identify an approach to distinguish legitimate profiles and fake profiles in LinkedIn, based only on limited publicly accessible profile information.

In order to carry out this project, we needed a set of fake profiles to serve as a comparison data set. Most current research in this area uses simulated fake profiles for this purpose. While this is a valid methodology, it has the potential to either over-emphasize expected characteristics of fake profiles, or omit characteristics that exist in real-world fake profiles. Ideally, we would like to work with a set of known fake LinkedIn profiles, which would avoid the potential issues associated with simulated data. We were able to find a set of web sources that provide lists of verified fake LinkedIn fake profiles that have been manually identified and checked, which allowed us to use verified fake profile data for our research.

In this research, we applied four data mining techniques, Neural Network (NN), Support Vector Machine (SVM), Principal Component Analysis (PCA) and Weighted Average (WA) for fake profile identification, in several combinations, to determine which combination of techniques produces the most accurate differentiation between fake and legitimate profiles. *Our results show that our approach performs with an accuracy of 84% and false negative of 2.44%. This is comparable to the results reported by existing research, where the results are based on much more expansive profile data than we consider in our research. Further, when compared to approaches using similar amounts and types of data, our method provides an improvement of approximately 14% in accuracy.*

The rest of the paper is organized as follows. Section 2 provides an overview of existing research related to LinkedIn, as well as prior research on fake profile identification. Section 3 describes the LinkedIn dataset and the process followed to collect the data. Section 4 describes our experimental method and results. Section 5 discusses the experimental results and their implications. Section 6 identifies the limitations of the study and discusses future directions for our research. Section 7 concludes our discussion of this study.

## 2 RELATED WORK

In this section, we provide an overview of existing research on the LinkedIn network and fake social network profile identification.

### 2.1 Research based on LinkedIn data

To date, little research has been carried out using LinkedIn as the primary data source. Hsieh, Tiwari [18] analysed LinkedIn profiles to understand the probability of connections between two people based on their organizational overlap. Xiang, Neville [19] considered interaction activity and similarity of user profiles to develop an unsupervised model to estimate friendship strength using proprietary data from LinkedIn.

### 2.2 Detecting fake profiles

Current research describes a number of different strategies that have been developed for identifying fake social network profiles. Much of the literature reports the application of these strategies applied in different contexts (e.g., in different social networks, or on different feature sets). Here, we describe a set of exemplars of these strategies.

*2.2.1 In LinkedIn*

Kontaxis, Polakis [11] proposed an approach for detecting cloned profiles based on LinkedIn data. While this is similar in spirit to our research, the actual research questions addressed are quite different. In the Kontaxis approach, the research study asks the following question: "Given a known legitimate profile on LinkedIn, can clones of that profile be accurately identified?" The authors propose a method that compares candidate clone profiles to a known legitimate profile to predict whether the target profile is a clone or not. In contrast, in our study here, we consider as input a LinkedIn profile whose status as fake or legitimate is unknown, and attempt to determine whether the profile is fake or legitimate, based solely on the publicly available content in the profile.

Kontaxis, Polakis [11] artificially induced fake cloned profile to demonstrate the efficacy of their approach. Additionally the number of fake profiles introduced is also just ten. Whereas in this research, we have relied on actual fake profiles reported in other sources and we have considered 34 such fake profiles.

### 2.2.2  *In other social networks*

Fire, Katz [3] used topology anomalies to identify spammers and fake profiles. They incorporated graph theory, supervised learning, parallel decision trees, and Naïve Bayes classifiers into their algorithm. Boshmaf, Muslukhov [20] adopted the traditional web-based Botnet design to build a group of adaptive social-bots as a socialbot network and analyzed its impact via millions of Facebook users. Jin, Takabi [10] analyzed the behavior of identity clone attacks and proposed a detection framework. Cao, Sirivianos [8] ranked users in online services to detect fake accounts. Their ranking algorithm is supported by social graphs according to the degree-normalized probability of a short random walk residing in the non-Sybil region. In a case study research, Krombholz, Merkl [7] analyzed privacy-related issues in social media contexts by creating desirable fake Facebook profiles and interacting with existing legitimate users, and documenting the information that could be harvested and analyzed from the users who interact with these fake profiles.

Chakraborty, Sundi [21] proposed an approach for detecting spam posts in Twitter using SVM. Lee, Eoff [4] developed a method to identify spam posters in Twitter using a Random Forest approach. Other past research [22, 23, 24] have also proposed approaches that rely on Twitter and Facebook posts, which are not available in Linkedin. Though the underlying tools used in all these studies are different, all heavily rely on actual twitter posts, i.e., on dynamic data. In the Linkedin case, the features for posting news and status are not commonly used by most Linkedin users (only 4% Linkedin users use Linkedin posting functionality [25], so any method that relies heavily on posting behavior cannot be directly used for fake profile detection in Linkedin.

Feizy, Wakeman [26] developed approach for determining fake profiles in MySpace based on profile data. However, they rely on more detailed data such as a profile's network connections and mutual friendships, which are not available in Linkedin unless the target user is directly connected. Additionally, Feizy, Wakeman [26] relied on fake profiles developed manually to demonstrate the efficacy of their approach. Obviously, generating artificial fake profiles creates the possibility of bias in the generation process, and does not demonstrate the capability of the proposed approach on real fake profiles. Similar to Feizy, Wakeman [26], Conti, Poovendran [27] developed an approach to detect fake Facebook profile using a graph-based network of Facebook friends. However, their approach also relies on artificially-generated fake profiles to demonstrate utility of the approach. Moreover, they assume that the fake profiles are within the network, and so more information (such as the target user's connection list) is available about these profiles.

In summary, our proposed approach differs in two dimensions from the existing approaches and thus provides a significant contribution in this stream of research. First, our approach identifies the minimum set of profile data to determine fakeness with high accuracy without requiring access to private data or dynamic posting behavior. Second, our approach considers verified fake profiles from existing known sources, rather then simulated profiles. We have used these fake profiles to demonstrate the applicability and accuracy of our approach.

# 3    LINKEDIN DATASET

For this study, we identified a data set consisting of both fake and legitimate LinkedIn profiles. In this section, we describe the process we followed to build our experimental dataset, including both fake profiles and legitimate profiles.

We identified the fake profiles in our dataset based on the existing evidence online, using LinkedIn profiles that have been investigated and declared to be fake by different sources in the internet. To identify such fake profiles, we searched blogs and web sites, and we were able to collect 34 fake Linkedin profiles. (Some profiles were identified by several sources.) The details of these fake profiles, along with their respective sources, are listed in the Appendix 1. This selection process is based on entirely on known available fake profiles, and we included all of the fake profiles that we could find in the dataset.

To identify legitimate profiles, we randomly selected 40 people affiliated with a major international university in Asia, and asked each of them to provide the public LinkedIn URL for a known legitimate profile in their network. Through this process, we collected 40 legitimate LinkedIn profiles. This process ensured randomness in the selection of these profiles.

For all profiles, we collected only the public URL, and did not create any connection link with these profiles. This ensures that we have access only to publicly available information.

Table 1 lists all the profile features we were able to capture publicly for both the legitimate and fake profiles, along with the maximum and average values of each profile feature across all profiles in our dataset. Due to LinkedIn restrictions, invariant of the number of connections or skills a profile might contain, the number of connections available publicly is limited to 500 and the number of skills available publicly is limited to 50. Therefore, rather than computing the normalized values via mean and standard deviation, we normalized the values by the maximum and minimum value of each feature. Since each profile feature value is absent at least once in either a legitimate or a fake profile, the minimum value for all features is 0. The maximum and average values are shown in Table 1.

| Profile feature | Maximum value | Average value | Description |
| --- | --- | --- | --- |
| No_Languages | 5 | 0.347 | Number of languages spoken |
| Profile_Summary | 1 | 0.52 | Presence of profile summary |
| No_Edu_Qualification | 7 | 1.467 | Number of education qualifications attained |
| No_Connections | 500 | 294.867 | Number of connections to other profiles |
| No_Recommendation | 37 | 2 | Number of recommendations made |
| Web_Site_URL | 1 | 0.28 | Presence of a URL for personal web site |
| No_Skills | 50 | 10.213 | Number of skills and expertise listed |
| No_Professions | 16 | 3.08 | Number of past and present professions listed |
| Profile_Image | 1 | 0.76 | Presence of a profile image |
| No_Awards | 10 | 0.56 | Number of awards won |
| Interests | 1 | 0.267 | Presence of any type of interests |
| No_LinkedIn_Groups | 51 | 8.907 | Number of LinkedIn groups and associations added |
| No_Publications | 16 | 0.613 | Number of publications listed |
| No_Projects | 7 | 0.24 | Number of work projects listed |
| No_Certificates | 9 | 0.267 | Number of certificates held |

*Table 1: Details of the profile features*

For analysis purposes, we developed three data sets (Dataset 1, Dataset 2, Dataset 3), each consisting of all 74 profiles in the overall LinkedIn data set. Within each data set, we randomly assigned half of the legitimate profiles (20 profiles) and half of the fake profiles (17 profiles) as a training data set, and assigned the other half as a test data set.

To demonstrate that the three datasets are truly randomly selected, we present the statistical significance of the each test dataset with respect to complete data in Table 2.

The statistical significance test is carried out using two sample tests along with Levene's Test for equality of variances. When the *Sig*-value of Levene's test is *>0.05* [28], we conclude that the mean and variance of two groups are the same, and thus the dataset is true representative of the complete data. As shown in Table 2, each feature has a higher *Sig* value (*>0.05*) across all three datasets. Thus, each dataset is a representation of the complete set at 5% significance level.

|  | Dataset 1 | | Dataset 2 | | Dataset 3 | |
|---|---|---|---|---|---|---|
|  | t | Sig. | t | Sig. | t | Sig. |
| No_Languages | 1.128 | .262 | -.993 | .323 | -.290 | .773 |
| Profile_Summary | -.347 | .729 | .400 | .690 | .531 | .597 |
| No_Edu_Qualifications | -.972 | .333 | 1.194 | .235 | 1.652 | .109 |
| No_Connections | 1.113 | .268 | -1.289 | .200 | -.066 | .948 |
| No_Recommendations | .147 | .883 | -.161 | .872 | .056 | .955 |
| Web_Site_URL | -.639 | .524 | .774 | .441 | -.063 | .950 |
| No_Skills | .575 | .566 | -.629 | .531 | .345 | .731 |
| No_Professions | -.138 | .891 | .164 | .870 | -.029 | .977 |
| Profile_Image | .294 | .769 | -.348 | .728 | .228 | .820 |
| No_Awards | -.415 | .679 | .504 | .615 | -.521 | .604 |
| Interests | .151 | .880 | -.172 | .864 | -.073 | .942 |
| No_LinkedIn_Groups | .093 | .926 | -.098 | .922 | .449 | .654 |
| No_Publications | -.134 | .894 | .157 | .876 | .125 | .901 |
| No_Projects | 1.809 | .074 | -1.074 | .285 | -.182 | .856 |
| No_Certificates | .895 | .373 | -.828 | .410 | .861 | .391 |
| Legitimacy | .289 | .773 | -.335 | .738 | -.121 | .904 |

*Table 2: Statistical significance of datasets*

## 4 METHOD AND RESULTS

In this research, we applied four well-known data mining techniques, Neural network (NN), Support vector machine (SVM), Principal component analysis (PCA) and Weighted average (WA) to the problem of fake profile identification in LinkedIn. Typically, social network research (including spam message identification, profile cloning and intruder detection) employs NN and SVM as the principal mining techniques. PCA is generally applied to reduce the number of dimensions of the datasets [29], and WA is used as a supportive implementation for optimizing results in the other data mining techniques [30, 31]. While we build on current methods in this work, we use WA as an independent solution to classify the data, rather than as a method for optimizing results.

In this section, we describe our proposed approach. Specifically, we explain how each technique is used in the process of data mining to differentiate legitimate and fake profiles. The process has three main mining steps. In the first step, we extract profile features using PCA. In the second step, we use three independent methods, NN, SVM and WA, to predict fake and legitimate profiles. NN and SVM follow a training-testing approach, while WA uses feature weights to generate a profile index for prediction. In the third step, we calculate accuracy rates across NN, SVM, and WA to compare the techniques (see Figure 2). For this research, we define accuracy as follows:

$$\% \ Accuracy = \frac{Total \ number \ of \ correctly \ identified \ profiles, both \ fake \ or \ legitimate}{Total \ number \ of \ profiles} \times 100$$

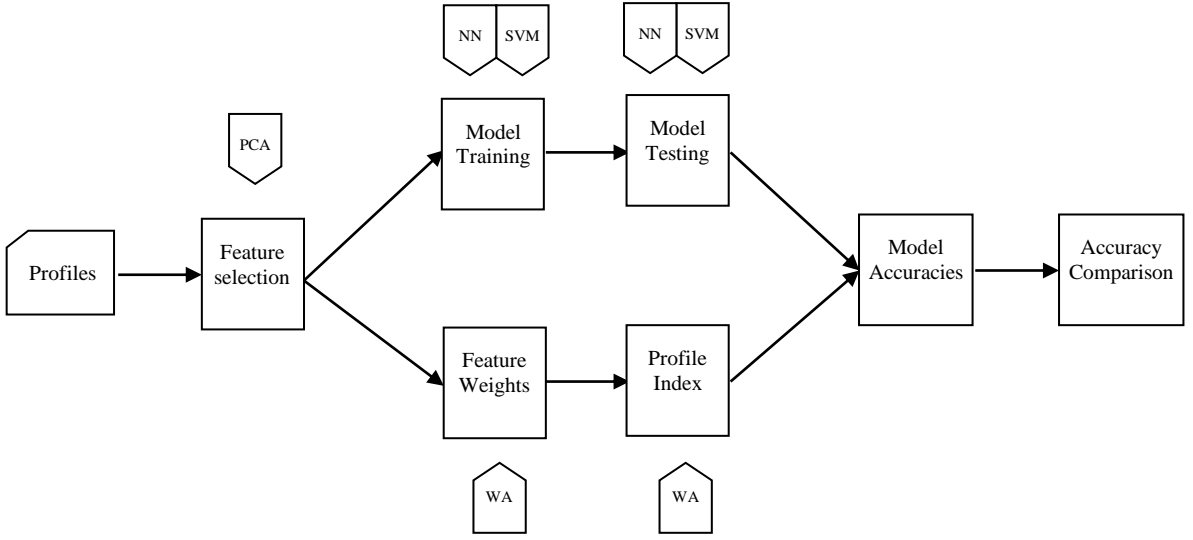

*Figure 2: Design approach to calculate accuracy rates for data mining techniques*

Moreover, for each analysis True Positive Rate (TPR) and True Negative Rate (TNR) were calculated through the false positive and false negative values. We define TPR and TNR as follows, where the term "false positive" refers to the number of fake profiles identified as legitimate profiles, and the term "false negative" refers to the number of legitimate profiles identified as fake profiles:

$$TPR = \frac{True\ Positives}{(True\ Positives + False\ Negatives)}$$

$$TNR = \frac{True\ Negatives}{(True\ Negatives + False\ Positives)}$$

In the remainder of this section, we first describe the feature selection process using PCA, which determines which features will be used as input to the NN, SVM, and WA mining methods. We then describe the mining process using each of the mining methods (NN, SVM, and WA), and finally present the accuracy results for each method, both with and without PCA feature selection. Table 3 describes the major notation used in this work.

| | |
|---|---|
| $P$ | Profile set |
| $p$ | Index for profile $p = 1, \ldots, |P|$ |
| $F$ | Feature set |
| $f$ | Index for feature $f = 1, \ldots, |F|$ |
| $R$ | Set of selected principal components |
| $S_{fr}$ | Score of feature $f$ in the principal component $r \in R$ |
| $v_{fp}$ | A feature value of a feature $f$ in a profile $p \in P$ |
| $V_f$ | Total feature value of a feature $f$ for all profiles in $P$ |
| $A_f$ | Average value of a feature $f$ across all profiles in $P$ |
| $c_{fp}$ | Binary indicative variable indicating whether feature $f$ is present in profile $p$ |
| $C_f$ | Number of profiles in *P* where feature *f* is present |
| $w_f$ | Weight of feature *f* |
| $N_f$ | Normalized feature count of a feature *f* |
| $W_p$ | Weight of a profile *p* |

*Table 3: Notation for symbols*

## 4.1 Feature selection through Principal Component Analysis

In this research study, we use PCA for dimensionality reduction, i.e., to determine which profile features best explain the variance in the data set, and therefore should be considered in the data mining process. While there are a number of different mathematical methods for deriving PCA results, we used variance maximization, one of the simplest PCA methods.

In variance maximization, we successively identify the first principal component, i.e., the component that has the highest projection variance, which accounts for the greatest variance across the data set. We then consider the remaining variance to identify the next principal component. We continue this process for all features in the data set.

To ensure the sampling adequacy, we tested for Kaiser–Meyer–Olkin (KMO) and Bartlett's Test. The resulting KMO value was 0.724, which is higher than the acceptable level of 0.5. The results of Barlett's test are significant at $p<0.05$ [32]. This verifies that there are sufficient samples to proceed with the study.

We considered both fake and legitimate profiles in the process of calculating the variance score for each of the features. We used Eigen-decomposition, the most commonly-used calculation method for PCA, to find the number of components. Eigenvalues provide information about the variability in the data. We estimated the variation of the components and selected the components with Eigenvalues greater than 1 [33]. There we found 5 components, which account for 64.82% of the total variance. The component variations and their Eigenvalues are shown in Table 4.

We then checked each component feature score, which provides information about the structure of the observations, to identify features that either load into several components with a score value greater than 0.5, or do not load into any component with a score value greater than 0.5 [33]. In order to better understand the relationship between these features and extracted principal components, we used Varimax rotation to reload the features into the components, and found that some features still load into several components without acceptable score values. We iteratively removed these features and re-ran the PCA based feature selection process until there were no more features to remove. This approach for feature reduction is described in algorithmic format in Algorithm 1.

|           | Initial Eigenvalues |              |              |
|-----------|--------|------------|--------------|
| Component | Total  | % of Variance | Cumulative % |
| 1         | 4.375  | 29.170     | 29.170       |
| 2         | 1.801  | 12.008     | 41.177       |
| 3         | 1.290  | 8.601      | 49.778       |
| 4         | 1.195  | 7.970      | 57.748       |
| 5         | 1.061  | 7.075      | 64.823       |
| 6         | .934   | 6.225      | 71.049       |
| 7         | .840   | 5.603      | 76.652       |
| 8         | .820   | 5.464      | 82.116       |
| 9         | .618   | 4.119      | 86.234       |
| 10        | .522   | 3.479      | 89.714       |
| 11        | .440   | 2.936      | 92.650       |
| 12        | .369   | 2.462      | 95.111       |
| 13        | .284   | 1.891      | 97.002       |
| 14        | .242   | 1.611      | 98.613       |
| 15        | .208   | 1.387      | 100.000      |

*Table 4: Total variance explained by PCA*

Based on the results of Algorithm 1, we removed the following features: *Profile_Image*, *No_Awards*, *No_LinkedIn_Groups* and *No_Publications*. As shown in Table 5, the remaining features load into four components with a total variation of 66.15%. The KMO value is lower, 0.655, but is still higher than the recommended threshold (>0.5) with the same significance value.

---

**Algorithm 1: Feature reduction through PCA**
**Input:** $F$, set of all features
Initialize each $Z_f$ to zero, where $Z_f$ is an binary variable associated to feature $f$
Do
    Run *PCA* with *Varimax* rotation
    If (*Eigenvalue* $\geq 1$)
        Select $R$, where $R$ is the set of selected principal components
        Initialize $L$ to empty, where $L$ is a list
        For each $f \epsilon F$
            For each $r \epsilon R$
                If $S_{fr} > 0.5$, where $S_{fr}$ is feature scores for feature $f$
                    $Z_f = Z_f + 1$
            End For
            If $Z_f$ is not equal to *1*
                Add $f$ to $L$
            End If
        End For
    End If
    For Each $f \in L$
        Remove $f$ from $F$
While (*L length* > *0*)
**Output:** $F$ is the set of selected feature set

| Feature No. | Profile Feature | Component 1 | Component 2 | Component 3 | Component 4 |
|---|---|---|---|---|---|
| [1] | No_Languages | 0.614 | 0.098 | 0.218 | -0.162 |
| [2] | Profile_Summary | 0.623 | 0.016 | 0.247 | -0.097 |
| [3] | No_Edu_Qualifications | 0.827 | 0.139 | -0157 | 0.266 |
| [4] | No_Professions | 0.702 | 0.171 | 0.153 | 0.311 |
| [5] | Web_Site_URL | 0.195 | 0.860 | 0.106 | -0.046 |
| [6] | Interests | 0.079 | 0.913 | 0.025 | 0.040 |
| [7] | No_Connections | 0.208 | -.0177 | 0.684 | 0.157 |
| [8] | No_Recommendations | -0.007 | 0.161 | 0.800 | 0.076 |
| [9] | No_Skills | 0.426 | 0.335 | 0.695 | 0.122 |
| [10] | No_Projects | 0.164 | 0.072 | 0.154 | 0.685 |
| [11] | No_Certificates | -0.075 | -0.082 | 0.077 | 0.843 |

*Table 5: PCA-selected feature loadings*

To further clarify the independence across the selected features, we show the correlation matrix for the selected features in Table 6.

|   | [1] | [2] | [3] | [4] | [5] | [6] | [7] | [8] | [9] | [10] | [11] |
|---|---|---|---|---|---|---|---|---|---|---|---|
| [1] | 1.000 | | | | | | | | | | |
| [2] | 0.261 | 1.000 | | | | | | | | | |
| [3] | 0.311 | 0.379 | 1.000 | | | | | | | | |
| [4] | 0.287 | 0.296 | **0.624** | 1.000 | | | | | | | |
| [5] | 0.161 | 0.236 | 0.259 | 0.237 | 1.000 | | | | | | |
| [6] | 0.155 | 0.061 | 0.177 | 0.266 | **0.659** | 1.000 | | | | | |
| [7] | 0.095 | 0.169 | 0.105 | 0.340 | 0.036 | -0.053 | 1.000 | | | | |
| [8] | 0.120 | 0.251 | 0.021 | 0.205 | 0.161 | 0.141 | 0.278 | 1.000 | | | |
| [9] | 0.502 | 0.317 | 0.319 | 0.448 | 0.406 | 0.331 | 0.465 | 0.516 | 1.000 | | |
| [10] | 0.132 | 0.153 | 0.206 | 0.207 | 0.075 | 0.091 | 0.188 | 0.132 | 0.270 | 1.000 | |
| [11] | -0.065 | -0.034 | 0.123 | 0.156 | -0.090 | -0.054 | 0.101 | 0.136 | 0.120 | 0.319 | 1.000 |

*Table 6: Correlation matrix for selected features*

Table 6 shows that almost all of the correlations are less than 0.6. Two combinations (features [3] and [4], and features [5] and [6]) have values are marginally higher than 0.6. However, in both cases, the two features load into the same component: both feature [3] and feature [4] load into component 1, and features [5] and [6] load into component 2 (Table 3). Therefore, we can state that the selected features are not highly correlated to each another [34, 35]. We summarize the results of the PCA-based feature selection step in Table 6.

| Feature Name | Is Selected by PCA? ( Y – Yes, N – No) | Selected Feature Number |
|---|---|---|
| *No_Languages* | Y | [1] |
| *Profile_Summary* | Y | [2] |
| *No_Edu_Qualification* | Y | [3] |
| *No_Connections* | Y | [4] |
| *No_Recommendation* | Y | [5] |
| *Web_Site_URL* | Y | [6] |
| *No_Skills* | Y | [7] |
| *No_Professions* | Y | [8] |
| *Profile_Image* | N | |

| *No_Awards* | N | |
| *Interests* | Y | [9] |
| *No_LinkedIn_Groups* | N | |
| *No_Publications* | N | |
| *No_Projects* | Y | [10] |
| *No_Certificates* | Y | [11] |

*Table 7: List of selected and all profile features*

**4.2   Neural Network mining**

Currently, there are many Neural Network (NN) algorithms designed to train models through supervised learning or unsupervised learning. In this research, we are interested in supervised learning, where the selected profile features (Table 7) serve as input, and predicted legitimacy is the response variable. We selected the Resilient backpropagation (Rprop) algorithm as the base algorithm for NN mining. Rprop does not account for the magnitude of the partial derivatives of the patterns (only the sign) and works independently on each weight [36]. Rprop is considered to be one of the fastest algorithms in data mining [37]. We selected the *neuralnet* package (an implementation of Rprop) in the R project for statistical computing [38]. *Neuralnet* is flexible, allowing for the selection of custom-choice of error-function, the number of covariates with response variables, and the number of hidden layers with hidden neurons.

Since the response variable (legitimacy of the profile) is binary (if legitimate, then the value is 1; if fake, the value is 0), we chose the logistic function (default) as the activation function of the training, and cross-entropy (err.fct="ce") as the error function. To ensure that the output is mapped by the activation function to the interval [0, 1], we defined linear.output as FALSE [38]. With this preparation, we trained the model by determining the number of hidden neurons and layers in relation to the optimized results. After several iterations, the best result (with the highest accuracy) is achieved with one hidden layer with two neurons.

First, we trained the model for all three datasets with all of the original features (i.e., including the features identified for removal in the PCA step) and saved their models in different variables. We then ran the same process for each dataset after removing the features identified for removal in the PCA step, and saved the models to different variables. Figure 3 shows an example of how covariate, response variables and hidden neurons are linked with calculated weights in the result model for Dataset 3 for the PCA-selected features.

We used the "compute" function of the library to predict results for new data based on the stored NN models. Since the compute function automatically redefines the NN structure only to calculate the output for arbitrary covariates, we could easily determine the predictions for the legitimacy of each of the test datasets, both with all features and with PCA-selected features. We then compared the predicted results with the actual legitimacy values (i.e., whether the profile is actually fake or legitimate) and calculated the accuracy for each dataset with all features and with PCA-selected features. These results are shown in Table 8.

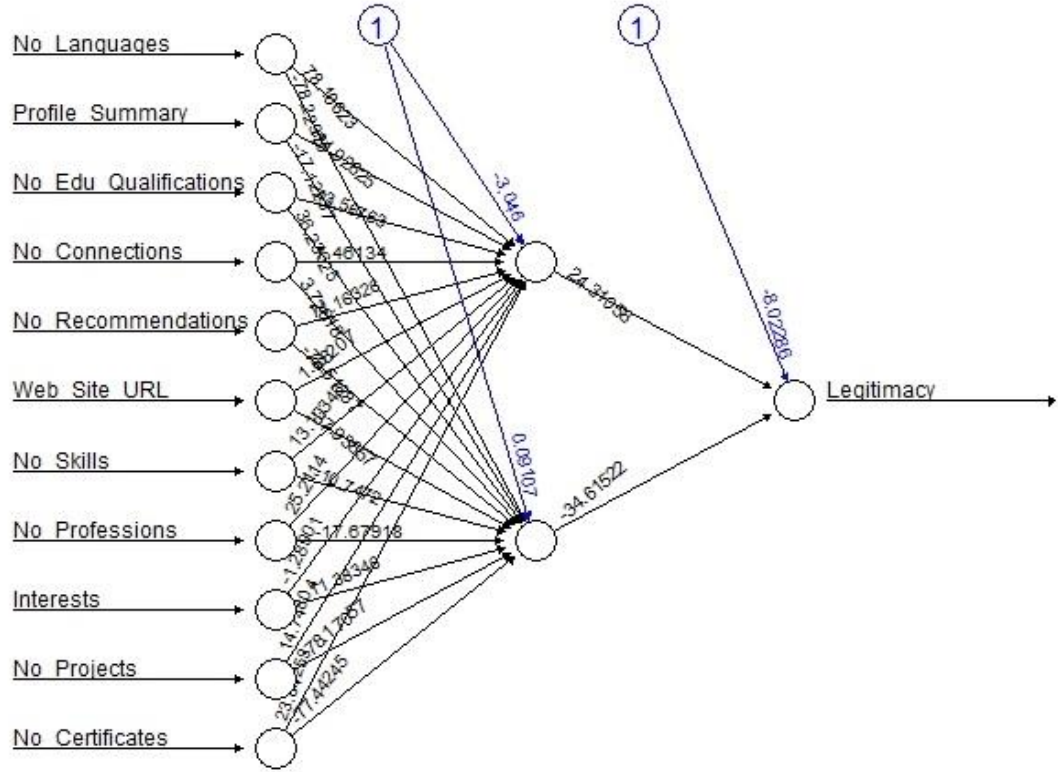

*Figure 3: NN plot for training model of Dataset 3 with PCA-selected features*

Table 8 shows that the TPR, TNR and accuracy results are higher for the PCA-selected features case as compared to the results when all features are used. When all features are considered, the model deteriorates due to the inclusion of unnecessary data points, which leads to over-fitting, and lower accuracy. This clearly demonstrates the importance of including the PCA step in our approach if NN is used for determining the legitimacy of a LinkedIn profile.

|  | **Dataset** | **Training error** | **TPR** | **TNR** | **Accuracy (%)** |
|---|---|---|---|---|---|
| All features | Dataset 1 | 0.043 | 0.79 | 0.93 | 84.85 |
|  | Dataset 2 | 0.083 | 0.62 | 0.75 | 68.29 |
|  | Dataset 3 | 0.064 | 0.74 | 0.92 | 86.11 |
|  | Average | 0.063 | 0.72 | 0.87 | 79.75 |
| Selected Features | Dataset 1 | 0.025 | 0.79 | 1.00 | 87.88 |
|  | Dataset 2 | 0.089 | 0.67 | 0.85 | 70.73 |
|  | Dataset 3 | 0.012 | 0.89 | 0.88 | 89.89 |
|  | Average | 0.042 | 0.78 | 0.91 | 82.83 |

*Table 8: Accuracy results obtained through Neural Network training*

## 4.3 Support Vector Machine mining

In this section, we describe how we applied a second supervised learning technique, a Support Vector Machine (SVM) based approach, to identify fake profiles. To develop an SVM training model, we applied C-support vector classification (C-svc) which is a Quadratical Programming (QP) solution. C-svc identifies the best possible hyperplane by measuring the margin between two classes using 2-norm of the normal vector and norm-1 is used for feature selection within the SVM method [39]. (This refers to feature selection within SVM, which is different from the PCA-based feature selection described in Section 4.1). According to Mercer's theorem [40], the kernel function K can be considered as equal to a dot-product in input space. Due to the nonlinearity of the profile features, SVM is able to create a random decision function in the input space on the kernel function.

We used both the Radial Basis function (RBF) kernel and Polynomial kernel as kernel functions in order to better understand the performance of SVM on our dataset.

We selected the RBF kernel because it uses the heuristic in sigest to calculate better sigma values, and we did not need to assign values to the kernel parameters. The Radial basis function kernel K can be written as

$$K(X_i, X_j) = e^{\gamma |X_i - X_j|^2}$$

We selected the Polynomial kernel because it uses combinations of features of the input sample instead of determining their similarity independently. The Polynomial kernel function can be written as,

$$K(X_i, X_j) = (-\gamma X_i \cdot X_j + C)^d$$

When C=0, the kernel is called homogenous.

For both kernels, $K(X_i, X_j) = \varphi(X_i) \cdot \varphi(X_j); \gamma = -\frac{1}{2\sigma^2}$

The transformation function $\varphi$ maps a dot product of input data points into a higher dimensional feature space where the non-linear patterns demonstrate linearity. γ is a parameter and γ >0.

We used KSVM (function of R, kernlab package), an implementation of a C-svc classifier, to train the SVM model. KSVM uses the Sequential Minimal Optimization (SMO) algorithm for solving the SVM quadratic programming (QP) optimization problem [41]. We generated the training models with the two proposed kernel functions (RBF and Polynomial) to create SVM models for all three training datasets, both with all features and with only PCA-selected features. We then tested the models using the test datasets. In this way we have total 12 models (2 Kernel functions, 3 datasets – each with both all features and PCA-selected features) to test and compare. We tested each with the appropriate test dataset, and calculated the Accuracy, TPR, and TNR rates. The consolidated results are presented in the Table 9.

|  |  | RBF kernel | | | Polynomial kernel | | |
|---|---|---|---|---|---|---|---|
|  | Dataset | TPR | TNR | Accuracy (%) | TPR | TNR | Accuracy (%) |
| All features | Dataset 1 | 0.63 | 0.87 | 78.79 | 0.79 | 0.92 | 84.85 |
| | Dataset 2 | 0.68 | 0.76 | 70.03 | 0.81 | 0.78 | 73.17 |
| | Dataset 3 | 0.79 | 0.92 | 88.89 | 0.89 | 1.00 | 91.67 |
| | Average | 0.70 | 0.85 | 79.24 | 0.83 | 0.90 | 83.23 |
| Selected features | Dataset 1 | 0.68 | 0.86 | 75.76 | 0.82 | 0.95 | 89.85 |
| | Dataset 2 | 0.75 | 0.85 | 78.05 | 0.81 | 0.87 | 75.61 |
| | Dataset 3 | 0.82 | 0.94 | 91.67 | 0.97 | 1.00 | 96.56 |
| | Average | 0.75 | 0.88 | 81.83 | 0.87 | 0.94 | 87.34 |

*Table 9: Accuracy results obtained through Support Vector Machine training*

In each of the scenarios, the Polynomial kernel delivered optimized results with fewer vectors in comparison to the Radial Basis Kernel. Since we need to compute the dot product of each support vector with the test point, the computational complexity of the model is linear to the number of support vectors.

Table 9 shows that the Polynomial Kernel performs better than the Radial Basis kernel in all cases, both with all features and with only PCA-selected features. For all three metrics, TPR, TNR and Accuracy, the case of the Polynomial Kernel with PCA-selected features provides the strongest results.

From Table 9 we also note that the TNR is higher than the TPR, indicating the approach is more successful in identifying a fake profile as a fake profile, than identifying legitimate profile as a legitimate profile. This is important in the LinkedIn scenario, which is the focus of our research, because false negatives carry greater risks than false positives. Consider an employer who pursues a potential candidate employee identified through LinkedIn, only to realize in middle or later stages of the process that the candidate's profile in Linkedin is fake. In this scenario, the organization has devoted resources in an unnecessary recruitment cycle, wasting the organization's valuable time and financial resources.

*Based on the above SVM discussion, we conclude that the Polynomial kernel applied on PCA-selected profile features provides the highest accuracy for identifying fake profiles, with the lowest percentage of false negatives across all SVM cases.*

### 4.4 Weighted Average mining

As a third mining alternative, we analyzed the dataset through a Weighted Average (WA) calculation of profile feature vectors, and then estimated the legitimate profiles index. During this process, we first derived the average value for each feature (normalized to a (0…1) range), based on all the legitimate profiles.

The total feature value can be computed as follows:

$$V_f = \sum_p v_{fp}$$

And feature average: $A_f = V_f/|P|$

We also count the number of times a feature is present in a legitimate profile as,

$$c_{fp} = \begin{cases} 1 \text{ if } v_{fp} > 0 \\ 0 \text{ if } v_{fp} = 0 \end{cases}$$

Then the total feature count for all the profiles is

$$C_f = \sum_p c_{fp}$$

For example, the average value for number of connections in a legitimate profile is 0.741 and the total feature count is 39. Table 10 shows the average value of each feature and its total count. Based on these two values we created a feature weight as:

$$w_f = A_f N_f$$

where, normalized feature count is,

$$N_f = \frac{C_f - max_f(C_f)}{max_f(C_f) - min_f(C_f)}$$

| Feature | Feature count | Feature average | Feature Weight |
|---|---|---|---|
| *No_Languages* | 15 | 0.135 | 0.037 |
| *Profile_summary* | 21 | 0.525 | 0.239 |
| *No_edu_qualifications* | 30 | 0.275 | 0.2 |
| *No_connections* | 39 | 0.741 | 0.741 |
| *No_recommendation* | 23 | 0.101 | 0.052 |
| *Web_site_url* | 15 | 0.375 | 0.102 |
| *No_Skills* | 30 | 0.389 | 0.283 |
| *No_professions* | 34 | 0.283 | 0.24 |
| *Profile_image* | 34 | 0.85 | 0.721 |
| *No_Awards* | 11 | 0.087 | 0.013 |
| *Interests* | 14 | 0.35 | 0.085 |
| *No_LinkedIn_Groups* | 25 | 0.232 | 0.133 |
| *No_Publications* | 8 | 0.072 | 0.004 |
| *No_projects* | 7 | 0.071 | 0.002 |
| *No_certificates* | 6 | 0.056 | 0 |

*Table 10: Feature count and weight for all features*

In the next step, we derived a profile weight $W_p$ (for both fake and legitimate profiles) for each profile for all three test datasets using the feature weights.

$$W_p = \sum_f w_f v_{fp}$$

We then estimated the profile index *I* for each dataset by averaging all the profile weights of a particular dataset. The profile index is,

$$I = \frac{\sum_p W_p}{|P|}$$

We calculated profile indices for each dataset for both all features and the PCA-selected features. We then compared the profile weight of each profile in each dataset with the respective profile indexes. If the profile weight is greater than the profile index, then we predict that the profile is legitimate; if it is less, then we predict that the profile is fake. Table 11 shows the accuracy of the results for each dataset for all features and PCA-selected features, along with the respective profile index for each case. The results in Table 11 demonstrate that WA performs better for the PCA-selected feature cases than the all features case for Dataset 1 and Dataset 3; however, for Dataset 2, WA performs much better for the all features case as compared to the PCA-selected feature case. We could not find any reason for this behaviour other than reduced reliability using WA for such detection. Similar to the SVM case, in this scenario TNR is higher than TPR, indicating the approach is more successful in identifying fake profiles as fake profiles than it is in identifying legitimate profiles as legitimate profiles, which matches with our use case scenario as described earlier.

|  | **Dataset** | **Profile Index** | **TPR** | **TNR** | **Accuracy** |
|---|---|---|---|---|---|
| All features | Dataset 1 | 1.43 | 0.79 | 0.71 | 75.76 |
|  | Dataset 2 | 1.28 | 0.76 | 0.85 | 80.49 |
|  | Dataset 3 | 1.41 | 0.68 | 0.74 | 72.22 |
|  | Average | 1.37 | 0.74 | 0.77 | 76.16 |
| Selected features | Dataset 1 | 0.84 | 0.79 | 0.93 | 84.85 |
|  | Dataset 2 | 0.73 | 0.61 | 0.69 | 63.41 |
|  | Dataset 3 | 0.82 | 0.71 | 0.79 | 75.79 |
|  | Average | 0.80 | 0.70 | 0.80 | 74.68 |

*Table 11: Accuracy results obtained through Weighted Average mining*

# 5   ANALYSIS OF RESULTS

One of the challenges of our research has been the collection of fake Linkedin profiles. Rather than relying on generated fake profiles, as has been done in past research, we rely on actual fake profiles reported in other sources. However, this resulted in a very small dataset, as the number of available verified fake profiles is not large. The small size of the dataset could lead a reader to question the validity of our approach. To address this concern, in this section we demonstrate the robustness of our approach by varying the dataset size and demonstrating the statistical significance of the result. Specifically, we demonstrate the robustness of our approach by demonstrating how TPR and TNR vary with differing training and test dataset sizes using the best-performing method we found, SVM with polynomial kernel using PCA-selected features.

To demonstrate the impact of training dataset size, we randomly select part of the training dataset for each of the three datasets, and measure the TPR and TNR using the corresponding full test dataset for each of the three datasets. We vary the portion of training data from 25% to 100%, and calculated TPR and TNR for each training data percentage for each dataset. Figure 4 shows the average TPR and TNR (averaged across all three datasets) as the percentage of training dataset varies. For lower values of training data percentage, both the TPR and TNR are relatively low, with accuracy similar to guessing whether a profile is fake or legitimate. Intuitively, this makes sense – the training dataset is very small, so one would expect relatively low accuracy in such a scenario. As the percentage of the dataset used for training grows, accuracy grows as well.  However, as the training dataset size is increased beyond 75%, both TPR and TNR stabilize. This indicates that beyond training dataset size of 75%, i.e., 28 training profiles, having a larger training dataset would only marginally improve the performance of our approach.

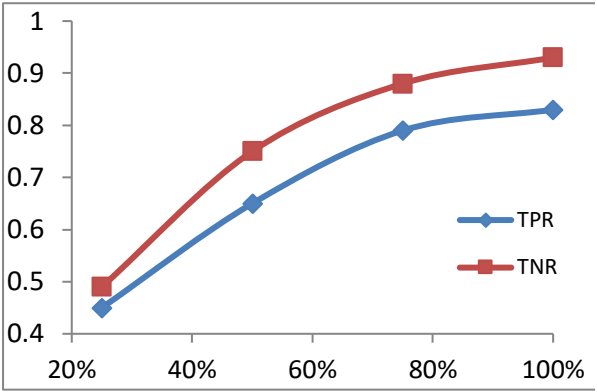 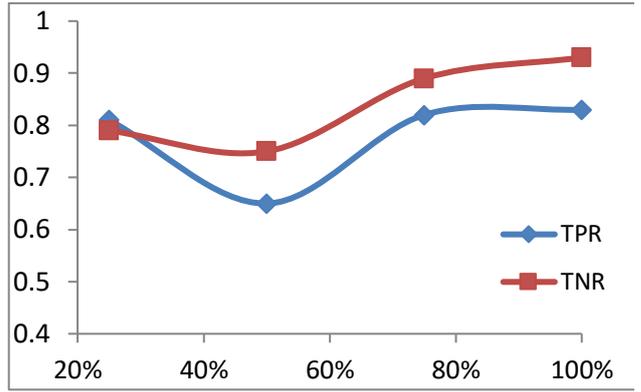

*Figure 4: Variation of TPR and TNR with training dataset size (testing dataset size = 100%)*   *Figure 5: Variation of TPR and TNR with testing dataset size (training dataset size = 100%)*

To demonstrate the stability of our results, we consider the impact of test dataset size. For each dataset, we build predictive models using the full-size training dataset (i.e., 37 profiles), and vary the percentage of test data considered from 25% to 100%. We plot how the average TPR and TNR vary with the percentage of test dataset tested in Figure 5. Figure 5 shows that lower percentages of test data considered (<=50%) provide TPR and TNR results that are not stable. However, at higher portions of test dataset size considered (between 75% and 100%), we see marginal differences in accuracy results.

To summarize, *we see an impact on accuracy results as the sizes of training and test datasets are varied; however, the impact is marginal at higher percentages of training and test dataset considered (>=75% for both training and test datasets) within the limited size of our data. This implies that the addition of further data points would likely provide only marginal increases in predictive accuracy.*

To demonstrate that the results of our model do not significantly differ from the actual results, we have computed the *p*-values through the McNemar test. We test the null hypothesis of "marginal homogeneity" - "there is no difference between the model identification and actual profile types". The opposite hypothesis is "there is a difference between model identification and actual profile types". However, since our ultimate goal is to demonstrate that the null hypothesis is true, i.e., the opposite hypothesis is false; we were looking for higher *p* values out of our test. Table 12 declares the *p*-values for each model according to each dataset.

|  | **Dataset 1** | **Dataset 2** | **Dataset 3** |
|---|---|---|---|
| **Support Vector Machine (Polynomial Kernel) with PCA selected features** | 0.3736 | 0.5050 | 0.2482 |

*Table 12: p-values of legitimacy determination vs. actual profile types*

Based on the results, we can see that for SVM with Polynomial Kernel and PCA selected features, even at a 90% confidence level, we cannot reject the null hypothesis. This implies that there is no significant difference in results between results of our approach and the actual profile types, and that *our approach is significantly effective in identifying the fake profiles*.

## 6    DISSCUSSION

In this research, we have compared the results of three data mining techniques to determine the most appropriate approach to differentiate legitimate profiles from fake profiles in LinkedIn. We first compare the results of our own study, comparing the accuracy of NN, SVM, and WA, both for all features and for only PCA-selected features. We then consider these accuracy results in the context of other research in fake social network profile identification.

Table 13 summarizes the final accuracy values for each technique as an average across all three datasets, as well as the average false positive rate and false negative rates for each technique.

|  | Feature selection | TPR | TNR | Accuracy (%) |
|---|---|---|---|---|
| **Neural Network** | All features | 0.72 | 0.87 | 79.75 |
|  | PCA-selected features | 0.78 | 0.91 | 82.83 |
| **Support Vector Machine (Polynomial Kernel)** | All features | 0.83 | 0.90 | 83.23 |
|  | PCA-selected features | 0.87 | 0.94 | 87.34 |
| **Weighted Average** | All features | 0.74 | 0.77 | 76.16 |
|  | PCA-selected features | 0.70 | 0.80 | 74.68 |

*Table 13: Accuracy comparison for the three techniques NN, SVM, and WA*

In Table 13, we show only the Polynomial kernel results for the SVM case. In our scenario, the Polynomial kernel provides greater accuracy compared to the RBF kernel, especially for false negatives in the PCA-selected features case (see Table 7). Therefore, we include only Polynomial kernel results for comparison with the NN and WA cases.

Based on the final accuracy rates, SVM clearly provides the highest accuracy rate among the three techniques, regardless of the number of features considered. However, the difference between the NN and SVM cases is 3.48% when all features are selected and 4.51% when only the PCA-selected features are considered. Based on the theoretical rationale [42, 43], SVM is preferred for the dataset we have, because SVM can compute results using fewer training data points than the other techniques, and does not suffer from local extrema. WA performs better with all features for dataset 3 (see Table 13), however overall it does not perform well compared to NN and SVM, and we remove WA from consideration as a technique for fake profile detection.

The TPR and TNR columns show the percentages of legitimate profiles detected as legitimate and percentage of fake profiles detected as fake, respectively. Compared to the false positive case, false negatives carry a greater risk, because the costs and risks are higher when a fake profile is identified as legitimate, as compared to the case when a legitimate profile is identified as fake. As shown in Table 13, SVM with selected feature has the highest TNR value (0.94). Thus, among all three approaches (NN, SVM and WA), *SVM with polynomial kernel gives the most accurate result, with the highest true negative rate (TNR)* for the task of fake profile identification in LinkedIn.

Among the three techniques considered, both NN and SVM provide higher accuracy when the features are selected through PCA. For both datasets 1 and 2, the accuracy values for the PCA-selected features are greater than the case where all features are considered. Further, the false negative value is lower for all three techniques when only PCA-selected features are used for legitimacy prediction. Thus, the PCA-based feature selection step is important in the process of identifying false profiles in LinkedIn.

From the above discussion, we conclude that *PCA-based feature selection followed by SVM modelling with the Polynomial Kernel is the preferred approach for identifying fake profiles on LinkedIn, where only a limited number of profile features are publicly available.*

Next, we show how our result compares with the results of previously proposed approaches. It is difficult to implement and run the proposed approaches in the literature on our dataset, because virtually all current approaches assume that the dataset is far less limited than publicly-available LinkedIn data. In Table 14, we present the accuracy results reported in previous research, along with the social media platform on which it was applied, and a summary of the data set requirements of the approach. In this table, the study reporting the results is shown in the far right column. For each study, an accuracy rate is provided for each combination of (1) type of features considered; (2) set of features considered, and (3) social network providing the data, as reported in the study. Accuracy rates shown in bold are based only on static data. All other accuracy results consider dynamic data.

Table 14 shows accuracy rates between 60.99% and 97% for prior studies focused on fake profile identification, where virtually all of these studies used generated fake profiles in their experimental studies. In contrast, in our study here, we considered actual profiles (both fake and legitimate). Despite the differences in the datasets considered in this study and past research, our approach, with average accuracy of 87.34% and a 0.94 TNR value is on par with or superior to existing work in the area.

In most prior studies concerned with fake profile identification, researchers considered user activities as an important criterion in determining the legitimacy of a profile. This includes all the dynamic information associated with a profile (e.g., number of posts, information about friends and their behaviours, etc.). In the LinkedIn case, this type of dynamic data is not publicly available due to the restrictive nature of the site's privacy policies. The sole study using only static data reports accuracy rates of 60.99% to 69.25%, which is substantially lower than the accuracy rates we report in this study. In addition, virtually all prior studies listed in Table 14 analyzed thousands of profiles, and introduced simulated fake profiles into the study, rather than considering known fake profiles. In contrast, in our approach, we considered actual fake profiles on LinkedIn, and our approach requires only the limited static profile data that is publicly available for LinkedIn profiles. *Considering these significant differences, our results of 87.34% accuracy and 0.94 TNR demonstrate that it is possible to achieve similar accuracy to existing approaches, using significantly less data, both in terms of profile count as well as features considered.*

| Technique used | Accuracy (%) | Feature Types | Features | Social network | Source |
|---|---|---|---|---|---|
| *Principal Component Analysis + Support Vector Machine* | ***87.34*** | *Static* | *No_Languages, Profile_Summary, No_Edu_Qualification, No_Connections, No_Recommendation, Web_Site_URL, No_Skills, No_Professions, Profile_Image No_Awards, Interests, No_LinkedIn_Groups, No_Publications, No_Projects, No_Certificates* | *Linkedin* | *(This research)* |
| Support Vector Machine | 78 | Dynamic and Static | profile age, presence of profile image, followers and friends count, posts/messages, details of tweets | Twitter | [21]. |
| Naïve Bayes | 67 | | | | |
| Random Forest | 98.42 | Dynamic and Static | screen name, user description, details of followers, details of tweets | Twitter | [4] |
| Decision Tree | **69.25** | Static | profile's content such as age, gender, location | MySpace | [26] |
| Rule Learner | **66.63** | | | | |
| Nearest Neighborhood | **67.05** | | | | |
| Naïve Bayes | **60.99** | | | | |
| Decision Tree | 86.10 | Dynamic | profile's connectivity, the amounts and types of interactions (mutual friends, relationships, in/out degree, similarity among friends) | | |
| Rule Learner | 85.89 | | | | |
| Nearest Neighborhood | 84.59 | | | | |
| Naïve Bayes | 78.03 | | | | |
| Twitter's detection algorithm | 89 | Dynamic | details of tweets and details of friends and followers | Twitter | [22] |
| Weka Classifier: Decorate | 88.98 | Dynamic and Static | user demographics, user contributed content, user activities, user connections | Twitter | [23] |
| Weka Classifier: LibSVM (SVM) | 83.09 | | | | |
| Weka Classifier: Random Forest algorithm | 94.5 | Dynamic | number of friends, friend requests, details of short text messages | Twitter | [24] |
| | 97 | Dynamic | notifications, private message, wall posts, and status updates | Facebook | |

*Table 14: Accuracy comparison for prior research on fake profile identification*

## 7   LIMITATIONS AND FUTURE WORK

One of the main limitations in this work lies in the verification of published fake profiles, as there may be situations where a source classifies a profile as a fake profile without proper evidence. Another limitation lies in the distinction among cloned profiles. In such cases, it is difficult to determine which profile is the original legitimate profile, and which profiles are the fake clones generated based on it. Second, when a cloning attack occurs on a profile, we cannot identify which profile is legitimate, and which is fake.

In future work, we intend to apply our approach to other social networks for generalizability, to determine whether similar accuracy levels can be attained exclusively based on limited static profile data. Further, we intend to delve deeper into the current study to extend our work here. Specifically, we believe that SVM accuracy can be improved by further analysing the kernel, and fine-tuning the kernel parameters and tolerance levels [44].

Further, in general, NN is more accurate when there are a larger number of data points, and we would expect better results for a larger set of profiles. We anticipate that there will be significant challenges in increasing the number of profiles in the study. While it is not difficult to identify legitimate profiles, it is particularly challenging to increase the set of fake profiles. Finally, we plan to look more closely at the WA method, to see whether we can improve the WA results by combining it with another method, e.g., Fuzzy sets, k-nearest neighbour, or moving average.

## 8   CONCLUSION

In this paper we propose an approach to identify fake profiles in LinkedIn based on limited, static profile data. We considered approaches based on NN, SVM (Radial and Polynomial kernels), and WA, both with and without PCA feature selection for each case. Our results show that SVM with Polynomial Kernel using PCA-selected features provides the highest accuracy across the tested methods, with the lowest percentage of false negatives.

Much of the existing research on fake profile detection assumes the availability of both dynamic and static data. Further, in most of these studies, the fake profiles were simulated for analysis purposes. To our knowledge, this is the first research focused on identifying fake profiles in LinkedIn using only static profile feature data (dynamic data is not accessible in LinkedIn), and studying an experimental dataset that consists of both verified fake profiles as well as verified legitimate profiles. *We demonstrate that with limited profile data, our approach can identify fake profile with 87.34% accuracy and a 94% True Negative Rate, which is comparable to the results obtained by other existing approaches based on larger data sets and significantly more varied profile information.*

# APPENDIX

List of fake profiles with the source references

| | First Name | Last Name | Location | Industry | Source |
|---|---|---|---|---|---|
| 1 | Pamela | May | Holtsville, New York (Greater New York City Area) | Human Resources | http://www.dikomci.com/post/37712291401/how-to-spot-a-fake-profile-on-linkedin-and-facebook |
| 2 | Anthony | Soprano | Greater New York City Area | Gambling & Casinos | http://www.integratedalliances.com/linkedin/how-to-spot-a-fake-linkedin-profile |
| 3 | Annmarie | Augustine | United States | | http://www.integratedalliances.com/linkedin/how-to-spot-a-fake-linkedin-profile |
| 4 | Brittany | Wilkey | Holtsville, New York | Human Resources | http://www.dikomci.com/post/37712291401/how-to-spot-a-fake-profile-on-linkedin-and-facebook |
| 5 | Monica | Patel | Holtsville, New York | Human Resources | http://www.dikomci.com/post/37712291401/how-to-spot-a-fake-profile-on-linkedin-and-facebook |
| 6 | Kristin | Ventura | Holtsville, New York | Human Resources | http://www.dikomci.com/post/37712291401/how-to-spot-a-fake-profile-on-linkedin-and-facebook |
| 7 | Christine | Curtiss | Holtsville, New York | Human Resources | http://www.dikomci.com/post/37712291401/how-to-spot-a-fake-profile-on-linkedin-and-facebook |
| 8 | Simryn | Grewal | Holtsville, New York | Human Resources | http://www.dikomci.com/post/37712291401/how-to-spot-a-fake-profile-on-linkedin-and-facebook |
| 9 | Karen Simms LION I Accept Inv.,karen_simms@journalist.com | Professional | Phoenix, Arizona Area | Banking | http://kschang.hubpages.com/hub/How-to-spot-fake-profiles-on-LinkedIn-and-other-social-networking-sites |
| 10 | Robin | Sage | Norfolk, Virginia | Compter & Network Security | http://www.computerworld.com/s/article/9179507/Fake_i_femme_fatale_i_shows_social_network_risks?pageNumber=1 |
| 11 | Jessica | Trot | Iran | Accounting | http://kschang.hubpages.com/hub/How-to-spot-fake-profiles-on-LinkedIn-and-other-social-networking-sites# |
| 12 | cherry | cole | United Kingdom | Capital Markets | http://kschang.hubpages.com/hub/How-to-spot-fake-profiles-on-LinkedIn-and-other-social-networking-sites#slide4941142 |
| 13 | Bonny | Andrew | Canada | Chemicals | http://kschang.hubpages.com/hub/How-to-spot-fake-profiles-on-LinkedIn-and-other-social-networking-sites# |
| 14 | Danielle | Baker | Greater New York City Area | Pharmaceuticals | http://booleanblackbelt.com/2013/03/linkedin-catfish-fake-profiles-real-people-or-fake-photos/ |
| 15 | Elizabeth | Rose | San Francisco Bay Area | Oil & Energy | http://booleanblackbelt.com/2013/03/linkedin-catfish-fake-profiles-real-people-or-fake-photos/ |
| 16 | Elizabeth | Obrien | Scottsdale, Arizona (Phoenix, Arizona Area) | Internet | http://booleanblackbelt.com/2013/03/linkedin-catfish-fake-profiles-real-people-or-fake-photos/ |
| 17 | Lola | Bader | United States | | http://booleanblackbelt.com/2013/03/linkedin-catfish-fake-profiles-real-people-or-fake-photos/ |
| 18 | J Walter | Thompson | Pittsfield, Massachusetts Area | Marketing and Advertising | https://www.facebook.com/media/set/?set=a.442669819090199.107405.236195939737589&type=3 |
| 19 | Kathy | Hill | Palo Alto, California (San Francisco Bay Area) | Computer Networking | https://www.facebook.com/media/set/?set=a.442669819090199.107405.236195939737589&type=3 |
| 20 | Tessy | Donna | Senegal | no | https://www.facebook.com/media/set/?set=a.442669819090199.107405.236195939737589&type=3 |
| 21 | Sulaiman | Al Fahim | Dubuque, Iowa Area | Real Estate | http://linkedin-superstar.com/ |
| 22 | Angelina | Jolie | San Francisco Bay Area | Entertainment | http://linkedin-superstar.com/ |
| 23 | Akilina | Stalin | Nashua, New Hampshire (Greater Boston Area) | Marketing and Advertising | http://www.slideshare.net/augustinefou/fake-profiles-on-linkedin |
| 24 | April | Pierce | Fresno, California Area | Information Technology and Services | http://www.slideshare.net/augustinefou/fake-profiles-on-linkedin |
| 25 | Lura | Burlingame | Tucson, Arizona Area | Hospital & Health Care | http://www.slideshare.net/augustinefou/fake-profiles-on-linkedin |
| 26 | Sandra | Morgan | Greater New Orleans Area | Animation | http://www.slideshare.net/augustinefou/fake-profiles-on-linkedin |
| 27 | Barbie | Jolly | Yakima, Washington , Amerika Serikat | Pharmaceuticals | http://www.slideshare.net/augustinefou/fake-profiles-on-linkedin |
| 28 | Susan | Wall | San Francisco Bay , Amerika Serikat | Computer Networking | http://www.slideshare.net/augustinefou/fake-profiles-on-linkedin |
| 29 | Vera | Knight | Greater San Diego Area | Facilities Services | http://www.slideshare.net/augustinefou/fake-profiles-on-linkedin |
| 30 | Lorraine | Hollingsworth | Charlotte, North Carolina Area | Alternative Medicine | http://www.slideshare.net/augustinefou/fake-profiles-on-linkedin |
| 31 | Laura | Hayes | Sacramento, California (Sacramento, California Area) | Information Technology and Services | http://www.slideshare.net/augustinefou/fake-profiles-on-linkedin |
| 32 | Jack | bauer | Copenhagen Area, Denmark | Design | http://blog.mxlab.eu/2009/04/15/wordpress-comments-lead-to-fake-company-profiles-on-linkedin/ |
| 33 | mavis | gwenda | Phoenix, Arizona Area | Health, Wellness and Fitness | http://www.globalrecruitingroundtable.com/2012/08/06/identifying-fake-linkedin-members/#.UjzZ2D-NBSM |
| 34 | Larry | Willeam | Young America, Minnesota (Greater Minneapolis-St. Paul Area) | Oil & Energy | http://www.debbie-carr.com/linked-in-has-a-fair-share-of-scammers-and-fake-profiles-too/ |